\documentclass[runningheads]{llncs}
\usepackage[T1]{fontenc}
\usepackage{amsfonts}
\usepackage{graphicx}
\usepackage{booktabs}
\usepackage{multirow}
\usepackage{tikz}
\usetikzlibrary{shapes.geometric, arrows.meta}

\begin{document}
\title{Frequency-Scale Saliency for Spectral Descriptor Analysis in 3D Shape Retrieval}
\titlerunning{Frequency-Scale Saliency for Spectral Descriptor Analysis}

\author{Jianru Shen\orcidID{0009-0000-3546-9616}}
\authorrunning{J. Shen}
\institute{University of Montana, Missoula, MT 59812, USA \\
\email{js258133@umconnect.umt.edu}}
\maketitle              
\begin{abstract}
Classical spectral descriptors such as the Heat Kernel Signature and Wave Kernel Signature are widely used for non-rigid 3D shape retrieval, yet their failure modes remain poorly understood. We present a frequency-scale saliency framework that audits these descriptors by quantifying the retrieval-level contribution of each descriptor scale interval through ablation. We introduce class spectral fingerprints to characterize category-level scale dependence, and show that descriptor similarity between class pairs is substantially correlated with retrieval failure, with a Spearman correlation of 0.479. Experiments on SHREC'11 demonstrate that short scales dominate retrieval performance while long scales are harmful, that HKS and WKS exhibit distinct scale dependence patterns, and that saliency-weighted retrieval improves mAP on hard categories by 0.156, with cross-fold and random-weight controls confirming that the gain is stable and not due to arbitrary reweighting.

\end{abstract}

\section{Introduction}

Spectral descriptors such as the Heat Kernel Signature (HKS) \cite{sun2009hks} and Wave Kernel Signature (WKS) \cite{aubry2011wks} are widely used for non-rigid 3D shape retrieval. Both descriptors are defined over a range of diffusion or energy scales, each capturing shape geometry at a different level of detail. Despite their broad adoption, little is known about which scale intervals actually drive retrieval performance and which are harmful or redundant.

Existing evaluations of spectral descriptors focus on aggregate retrieval metrics such as mAP, which obscure per-scale contributions. When a descriptor fails on geometrically similar categories, it is unclear whether the failure stems from an inherent limitation of the spectral representation or from reliance on uninformative scale intervals. This lack of diagnostic capability makes it difficult to understand, compare, or improve spectral descriptors in a principled way.

We address this gap by proposing a frequency-scale saliency framework that audits spectral descriptors through ablation-based analysis. Our method quantifies the retrieval-level contribution of each scale interval, introduces class spectral fingerprints to characterize category-level scale dependence, and shows that descriptor similarity between class pairs is substantially correlated with retrieval failure. The framework provides an interpretable diagnosis of spectral descriptor limitations that goes beyond aggregate performance scores.

The main contributions of this paper are:
\begin{itemize}
\item An ablation-based frequency-scale saliency method that quantifies the retrieval-level contribution of each diffusion-time or energy-scale interval for HKS and WKS.
\item Class spectral fingerprints derived from per-query AP drop, revealing that hard categories exhibit unstable and oscillating scale dependence.
\item Evidence that descriptor similarity between category pairs is substantially correlated with retrieval failure (Spearman $\rho = 0.479$, $p < 10^{-25}$), and a robustness analysis showing that saliency-weighted retrieval remains stable under cross-fold weighting while outperforming random scale reweighting.
\end{itemize}

\section{Related Work}

\paragraph{Spectral shape descriptors.}
HKS~\cite{sun2009hks} encodes the heat diffusion process at multiple time scales, capturing both local and global geometry. WKS~\cite{aubry2011wks} uses energy band filtering on the Laplace-Beltrami eigenfunctions to produce a more frequency-selective descriptor. Bronstein and Kokkinos~\cite{bronstein2010scale} proposed a scale-invariant variant of HKS. Rustamov~\cite{rustamov2007laplace} proposed Laplace-Beltrami eigenfunctions for deformation-invariant representation, and Litman and Bronstein~\cite{litman2014learning} extended spectral descriptors through learning. A comprehensive survey of shape similarity methods is provided in~\cite{biasotti2016recent}, and a thorough treatment of non-rigid shape analysis is given in~\cite{bronstein2008numerical}. Despite broad adoption, the scale-level discriminative behavior of these descriptors has not been systematically analyzed.

\paragraph{3D shape retrieval.}
Shape retrieval benchmarks~\cite{lian2011shrec} evaluate methods using mAP and precision-recall curves. Shape Google~\cite{bronstein2011shape} introduced bag-of-features approaches for invariant retrieval. Intrinsic shape context descriptors~\cite{kokkinos2012intrinsic} and local surface signatures~\cite{tombari2010unique} represent alternative approaches to shape description. Sparse descriptor reconstruction~\cite{wan2017full} and canonical-form-based methods~\cite{pickup2018evaluation} provide further evaluation frameworks for non-rigid retrieval. Most comparisons treat descriptors as black boxes, reporting overall performance without examining which components contribute to success or failure.

\paragraph{Interpretability in shape analysis.}
The ShapeDNA approach~\cite{reuter2006laplace} uses global spectral properties for shape identification, and Reuter et al.~\cite{reuter2009discrete} studied discrete Laplace-Beltrami operators for shape analysis. Levy~\cite{levy2006laplace} examined Laplace-Beltrami eigenfunctions from an algorithmic perspective. Functional maps~\cite{ovsjanikov2012fmap} and partial correspondence methods~\cite{rodola2017partial} have explored spectral representations for shape matching. Frequency-scale saliency as a diagnostic tool for retrieval has not been explored; our work fills this gap by providing a systematic ablation-based analysis of scale contributions in spectral descriptor retrieval, offering interpretable diagnostics that go beyond aggregate retrieval metrics.

\section{Background}

Given a triangulated mesh, we compute the cotangent Laplacian $L$ and mass matrix $M$~\cite{reuter2009discrete,levy2006laplace}, and solve the generalized eigenvalue problem $L\phi_k = \lambda_k M\phi_k$. We discard the zero eigenvalue corresponding to the constant eigenfunction and retain the first $K$ positive eigenpairs $(\lambda_1, \phi_1), \ldots, (\lambda_K, \phi_K)$ with $0 < \lambda_1 \leq \lambda_2 \leq \cdots \leq \lambda_K$.

The Heat Kernel Signature~\cite{sun2009hks} at vertex $x$ and time scale $t$ is defined as:
\begin{equation}
\mathrm{HKS}(x, t) = \sum_{k=1}^{K} e^{-\lambda_k t} \, \phi_k(x)^2
\end{equation}
Small values of $t$ capture fine local geometry; large values capture global structure.

The Wave Kernel Signature~\cite{aubry2011wks} at vertex $x$ and energy $e$ is defined as:
\begin{equation}
\mathrm{WKS}(x, e) = C_e^{-1} \sum_{k=1}^{K} \exp\!\left(-\frac{(e - \log \lambda_k)^2}{2\sigma^2}\right) \phi_k(x)^2,
\end{equation}
where $C_e = \sum_{k=1}^{K} \exp(-(e - \log \lambda_k)^2 / (2\sigma^2))$ is the normalization constant. Each energy level $e$ selects eigenfunctions whose frequencies are close to $e$, providing more frequency-selective shape description than HKS.

For HKS we sample $T$ log-spaced time scales in the range $[4\pi^2 / \lambda_K, \, 4\pi^2 / \lambda_1]$, following standard practice~\cite{sun2009hks}. For WKS we sample $T$ linearly spaced energy levels in $[\log \lambda_1, \, \log \lambda_K]$ with bandwidth $\sigma = 7(e_{\max} - e_{\min})/T$~\cite{aubry2011wks}. Per-vertex descriptors are aggregated into a global shape descriptor via mean and max pooling, yielding a $2T$-dimensional vector normalized to unit $\ell_2$ norm.

\section{Methodology}
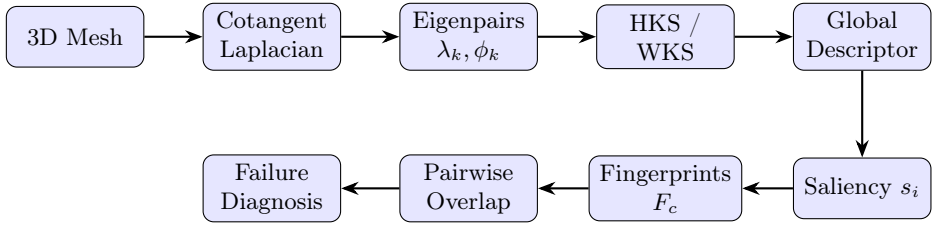
\begin{figure}[t]
\centering
\begin{tikzpicture}[
    box/.style={rectangle, rounded corners, draw=black, fill=blue!10,
                text width=1.6cm, align=center, minimum height=0.8cm, font=\small},
    arrow/.style={-{Stealth}, thick}
]
\node[box] (mesh) at (0,0)    {3D Mesh};
\node[box] (lap)  at (2.6,0)  {Cotangent Laplacian};
\node[box] (eig)  at (5.2,0)  {Eigenpairs $\lambda_k, \phi_k$};
\node[box] (desc) at (7.8,0)  {HKS / WKS};
\node[box] (agg)  at (10.4,0) {Global Descriptor};
\node[box] (sal)  at (10.4,-2){Saliency $s_i$};
\node[box, text width=1.8cm] (fp) at (7.8,-2) {Fingerprints $F_c$};
\node[box] (ovlp) at (5.2,-2) {Pairwise Overlap};
\node[box] (diag) at (2.6,-2) {Failure Diagnosis};

\draw[arrow] (mesh) -- (lap);
\draw[arrow] (lap)  -- (eig);
\draw[arrow] (eig)  -- (desc);
\draw[arrow] (desc) -- (agg);
\draw[arrow] (agg)  -- (sal);
\draw[arrow] (sal)  -- (fp);
\draw[arrow] (fp)   -- (ovlp);
\draw[arrow] (ovlp) -- (diag);
\end{tikzpicture}
\caption{Overview of the proposed frequency-scale saliency framework.}
\label{fig:pipeline}
\end{figure}

Given a mesh, we compute the cotangent Laplacian and truncate the first $K$ eigenpairs to construct per-vertex HKS and WKS descriptors, which are aggregated into global shape descriptors. We then apply ablation-based saliency analysis to identify discriminative scale intervals and derive class spectral fingerprints for failure diagnosis. Figure~\ref{fig:pipeline} provides an overview of the complete framework. All spectral descriptors are precomputed offline. Given the cached global descriptors, retrieval and saliency analysis require $O(N^2T)$ operations.

\subsection{Ablation-Based Frequency-Scale Saliency}
We define the saliency of scale interval $i$ as the drop in mAP when that interval is masked:
\begin{equation}
s_i = \mathrm{mAP}_{\mathrm{full}} - \mathrm{mAP}_{\mathrm{mask}_i}
\end{equation}
where $\mathrm{mask}_i$ denotes setting the $i$-th scale dimensions to zero in all descriptors before retrieval. Positive values indicate that interval $i$ contributes to retrieval; negative values indicate it is harmful.

\subsection{Class Spectral Fingerprints}
We define per-query saliency as the AP drop for a single query shape $x$:
\begin{equation}
s_i(x) = \mathrm{AP}(x)_{\mathrm{full}} - \mathrm{AP}(x)_{\mathrm{mask}_i}
\end{equation}
The class spectral fingerprint of category $c$ is the mean over all queries in that class:
\begin{equation}
F_c(i) = \mathbb{E}_{x \in c}[s_i(x)]
\end{equation}

\subsection{Failure Diagnosis via Descriptor Similarity}
For each pair of classes $(c_1, c_2)$, we compute the $\ell_2$ distance between normalized class mean descriptors~\cite{bronstein2011shape}. A low distance indicates that the two classes occupy similar regions of the descriptor space, making retrieval confusion structurally likely.

\subsection{Saliency-Weighted Retrieval}
Using the global saliency scores, we define a weighted distance~\cite{litman2014learning}:
\begin{equation}
d_w(x, y) = \sum_i \tilde{s}_i \, (x_i - y_i)^2, \quad \tilde{s}_i = \frac{\max(s_i, 0)}{\sum_j \max(s_j, 0)}
\end{equation}
This serves as a lightweight application of the saliency analysis rather than a primary contribution.

\section{Experiments}

\subsection{Dataset and Setup}

We evaluate on SHREC'11~\cite{lian2011shrec}, a benchmark of 600 watertight triangle meshes equally distributed across 30 categories. Meshes are loaded without additional processing and are directly used for cotangent Laplacian and lumped mass matrix computation. For each mesh, we truncate the first $K = 100$ eigenpairs and construct HKS descriptors over $T = 32$ log-spaced diffusion times and WKS descriptors over $T = 32$ linearly spaced energy levels. Per-vertex descriptors are aggregated via mean and max pooling into a 64-dimensional global descriptor, normalized to unit $\ell_2$ norm. All descriptors are precomputed and cached. We use leave-one-out retrieval: each shape serves as a query against the remaining 599 shapes. The primary metric is mean Average Precision (mAP), supplemented by Precision@5 and Top-1 accuracy.

\paragraph{Sensitivity to hyperparameters.}
Table~\ref{tab:sensitivity} reports mAP changes under varying numbers of eigenpairs $K$ and scale samples $T$. HKS performance remains stable across the evaluated settings. WKS is more sensitive to the number of sampled energy levels: performance degrades substantially at $T = 16$, while denser sampling at $T = 64$ can further improve retrieval accuracy. We use $K = 100$ and $T = 32$ as a practical baseline that balances descriptor dimensionality and retrieval performance.

\begin{table}[t]
\caption{Sensitivity analysis: mAP change relative to baseline ($K=100$, $T=32$) on SHREC'11.}\label{tab:sensitivity}
\centering
\begin{tabular}{ccccc}
\toprule
& & \multicolumn{2}{c}{$\Delta$ mAP} \\
\cmidrule(lr){3-4}
$K$ & $T$ & HKS & WKS \\
\midrule
50  & 16 & $-$0.048 & $-$0.165 \\
50  & 32 & $-$0.044 & $+$0.029 \\
50  & 64 & $-$0.045 & $+$0.077 \\
\midrule
100 & 16 & $-$0.001 & $-$0.272 \\
100 & 32 & $\pm$0.000 & $\pm$0.000 \\
100 & 64 & $+$0.000 & $+$0.089 \\
\midrule
150 & 16 & $-$0.009 & $-$0.361 \\
150 & 32 & $-$0.008 & $-$0.044 \\
150 & 64 & $-$0.008 & $+$0.076 \\
\bottomrule
\end{tabular}
\end{table}

\subsection{Baseline Retrieval}

Table~\ref{tab:retrieval} reports retrieval performance for HKS, WKS, their concatenation, and saliency-weighted WKS. WKS outperforms HKS by a substantial margin (0.810 vs. 0.714), and concatenating the two descriptors does not improve over WKS alone, suggesting that HKS introduces noise rather than complementary information. Saliency-weighted WKS achieves the highest overall mAP of 0.867.

\begin{table}[t]
\caption{Retrieval performance on SHREC'11. P@5 denotes Precision at 5.}\label{tab:retrieval}
\centering
\begin{tabular}{lccc}
\toprule
Method & mAP & P@5 & Top-1 \\
\midrule
HKS & 0.714 & 0.855 & 0.862 \\
WKS & 0.810 & 0.930 & 0.962 \\
HKS + WKS & 0.747 & 0.893 & 0.935 \\
WKS saliency-weighted & \textbf{0.867} & \textbf{0.939} & \textbf{0.963} \\
\bottomrule
\end{tabular}
\end{table}

\begin{figure}[t]
\includegraphics[width=0.75\textwidth]{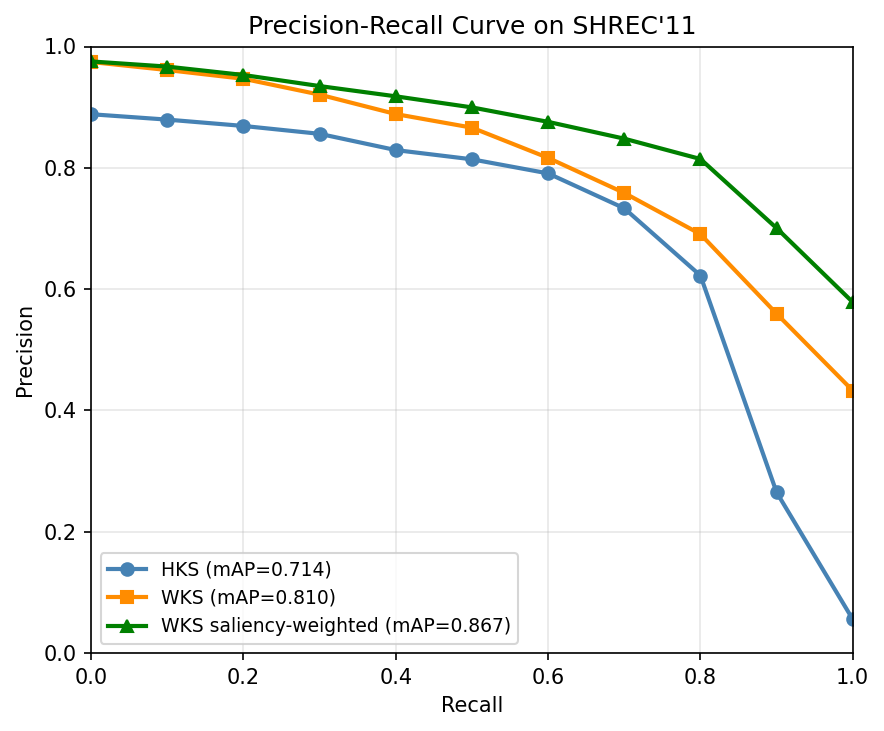}
\centering
\caption{Precision-recall curves on SHREC'11 for HKS, WKS, and saliency-weighted WKS.}
\label{fig:pr}
\end{figure}

Figure~\ref{fig:pr} shows the precision-recall curves for all methods. WKS saliency-weighted consistently achieves higher precision across all recall levels, with the most pronounced gap at high recall values where hard categories dominate.

\subsection{Frequency-Scale Saliency}
\label{sec:saliency}

\begin{figure}[t]
\includegraphics[width=\textwidth]{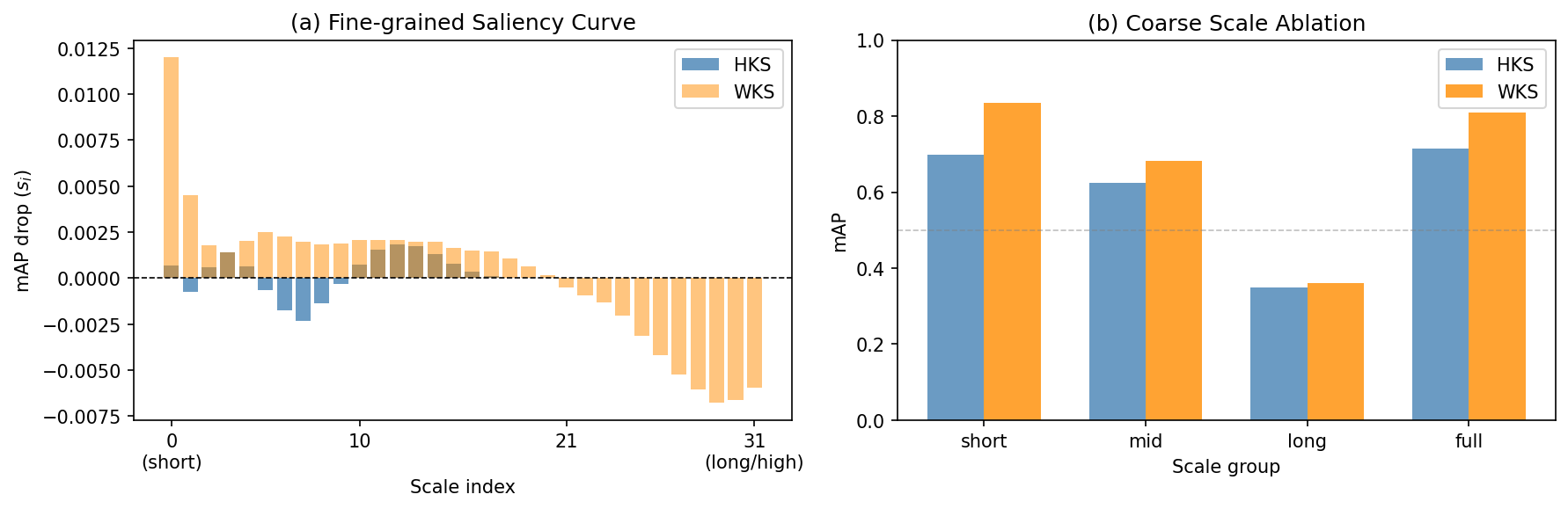}
\caption{(a) Fine-grained saliency curves for HKS and WKS. (b) Coarse ablation: mAP when only short, mid, or long scale intervals are retained.}
\label{fig:saliency}
\end{figure}

Figure~\ref{fig:saliency} shows the ablation-based saliency curves for HKS and WKS alongside coarse-scale ablation results. For both descriptors, short scales contribute most to retrieval performance, while long scales are consistently harmful. WKS concentrates discriminative power at the lowest energy scales, whereas HKS shows a broader mid-range dependence with notable negative contributions at short scales. This difference in scale dependence confirms that HKS and WKS rely on structurally distinct parts of the spectrum.

Table~\ref{tab:ablation} reports retrieval metrics for each scale group. We define the short, mid, and long groups as scale indices 0--10, 11--21, and 22--31, respectively, corresponding to a roughly equal three-way partition of the $T = 32$ sampled descriptor dimensions. Short scales alone recover most of the full descriptor performance for both HKS and WKS, while long scales perform substantially worse than the corresponding full descriptors. For WKS, short scales achieve a mAP of 0.836 compared to 0.810 for the full descriptor, suggesting that the remaining scales introduce noise.

\begin{table}[t]
\caption{Scale group ablation on SHREC'11. Each row uses only the specified scale interval.}\label{tab:ablation}
\centering
\begin{tabular}{llccc}
\toprule
Descriptor & Scale group & mAP & P@5 & Top-1 \\
\midrule
HKS & short & 0.699 & 0.844 & 0.860 \\
HKS & mid   & 0.625 & 0.784 & 0.823 \\
HKS & long  & 0.348 & 0.490 & 0.565 \\
HKS & full  & 0.714 & 0.855 & 0.862 \\
\midrule
WKS & short & \textbf{0.836} & \textbf{0.926} & \textbf{0.958} \\
WKS & mid   & 0.683 & 0.833 & 0.917 \\
WKS & long  & 0.362 & 0.466 & 0.570 \\
WKS & full  & 0.810 & 0.930 & 0.962 \\
\bottomrule
\end{tabular}
\end{table}

\subsection{Class Spectral Fingerprints}

\begin{figure}[t]
\includegraphics[width=\textwidth]{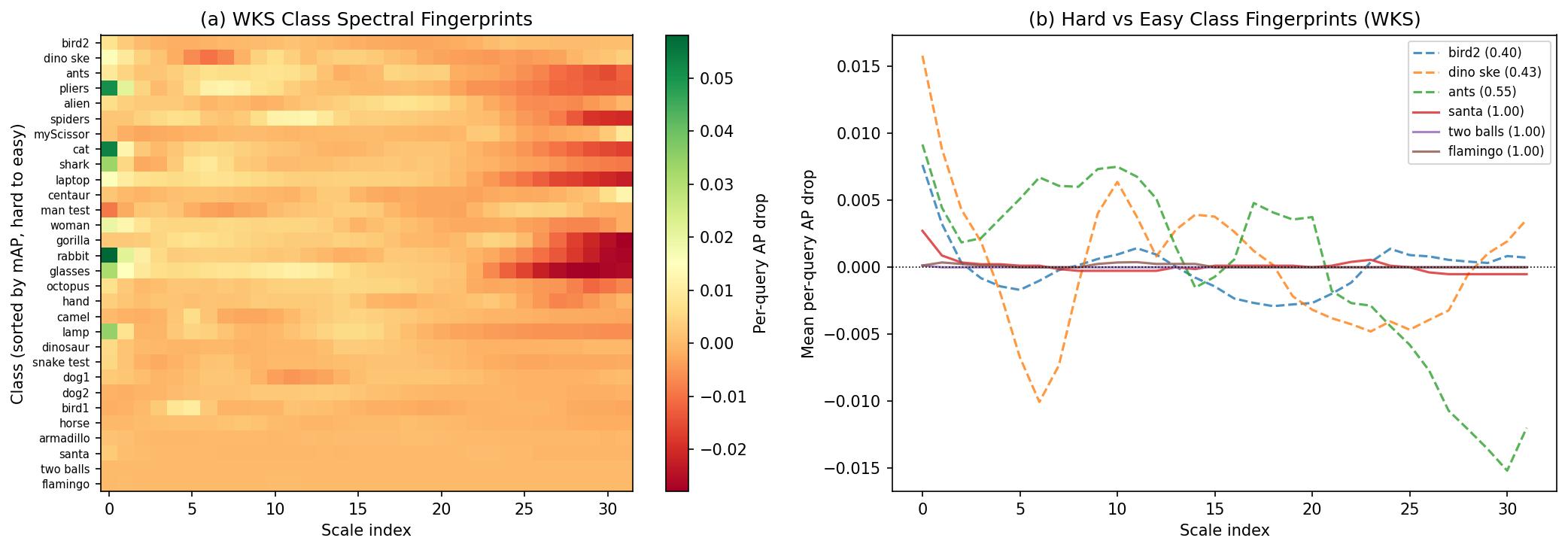}
\caption{(a) WKS class spectral fingerprints for all 30 categories, sorted by mAP from hardest to easiest. (b) Fingerprint curves for the three hardest and three easiest classes.}
\label{fig:fingerprints}
\end{figure}

Figure~\ref{fig:fingerprints} shows WKS class spectral fingerprints for all 30 categories, sorted by mAP from hardest to easiest. Easy classes such as \textit{flamingo} and \textit{two balls} exhibit flat, near-zero fingerprints, indicating stable and consistent scale dependence across shapes. Hard classes such as \textit{bird2} and \textit{dino ske} show irregular, oscillating profiles with large positive and negative contributions across scales, indicating that no single scale interval reliably discriminates these categories. This contrast suggests that fingerprint shape is a meaningful indicator of per-class retrieval difficulty.

\subsection{Fingerprint Overlap and Failure Diagnosis}

\begin{figure}[t]
\includegraphics[width=\textwidth]{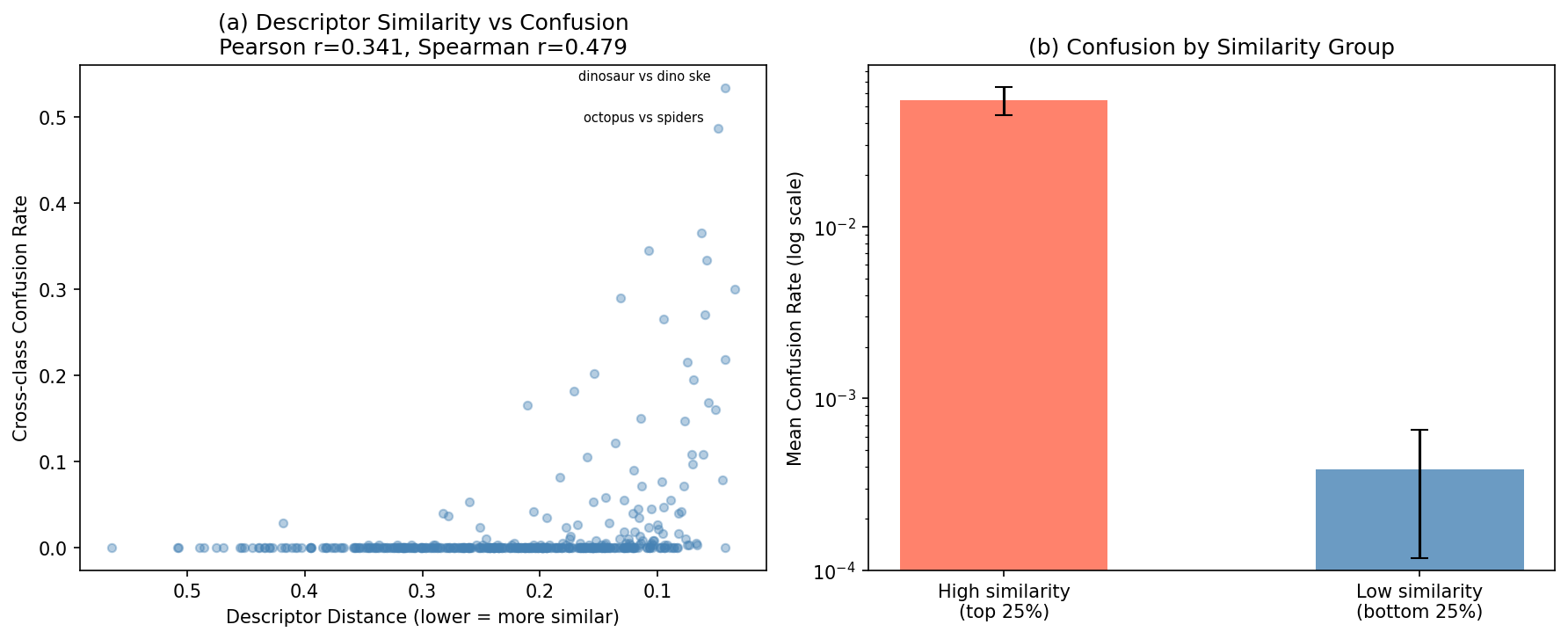}
\caption{(a) Descriptor similarity vs. cross-class confusion rate across 435 class pairs. (b) Mean confusion rate for high-similarity and low-similarity groups.}
\label{fig:overlap}
\end{figure}

Figure~\ref{fig:overlap} shows the relationship between descriptor similarity and cross-class confusion rate across all 435 class pairs. Pairs with higher descriptor similarity exhibit substantially higher confusion rates. Pearson correlation is 0.341 and Spearman correlation is 0.479, both significant at $p < 10^{-12}$. Grouping pairs by descriptor distance confirms this: the top-25\% most similar pairs yield a mean confusion rate of 0.055, compared to 0.0004 for the bottom-25\%, a difference of two orders of magnitude. Table~\ref{tab:hardpairs} lists the most confused category pairs, all of which share low descriptor distance. Both correlations are highly significant ($p < 10^{-12}$ for Pearson, $p < 10^{-25}$ for Spearman), confirming that the association is not due to chance.

\begin{table}[t]
\caption{Top confused category pairs ranked by confusion rate.}\label{tab:hardpairs}
\centering
\begin{tabular}{llcc}
\toprule
Class A & Class B & Confusion & Distance \\
\midrule
dinosaur & dino ske & 0.534 & 0.043 \\
octopus & spiders & 0.487 & 0.048 \\
ants & spiders & 0.366 & 0.063 \\
glasses & rabbit & 0.334 & 0.058 \\
\bottomrule
\end{tabular}
\end{table}

\begin{figure}[t]
\centering
\includegraphics[width=0.85\textwidth]{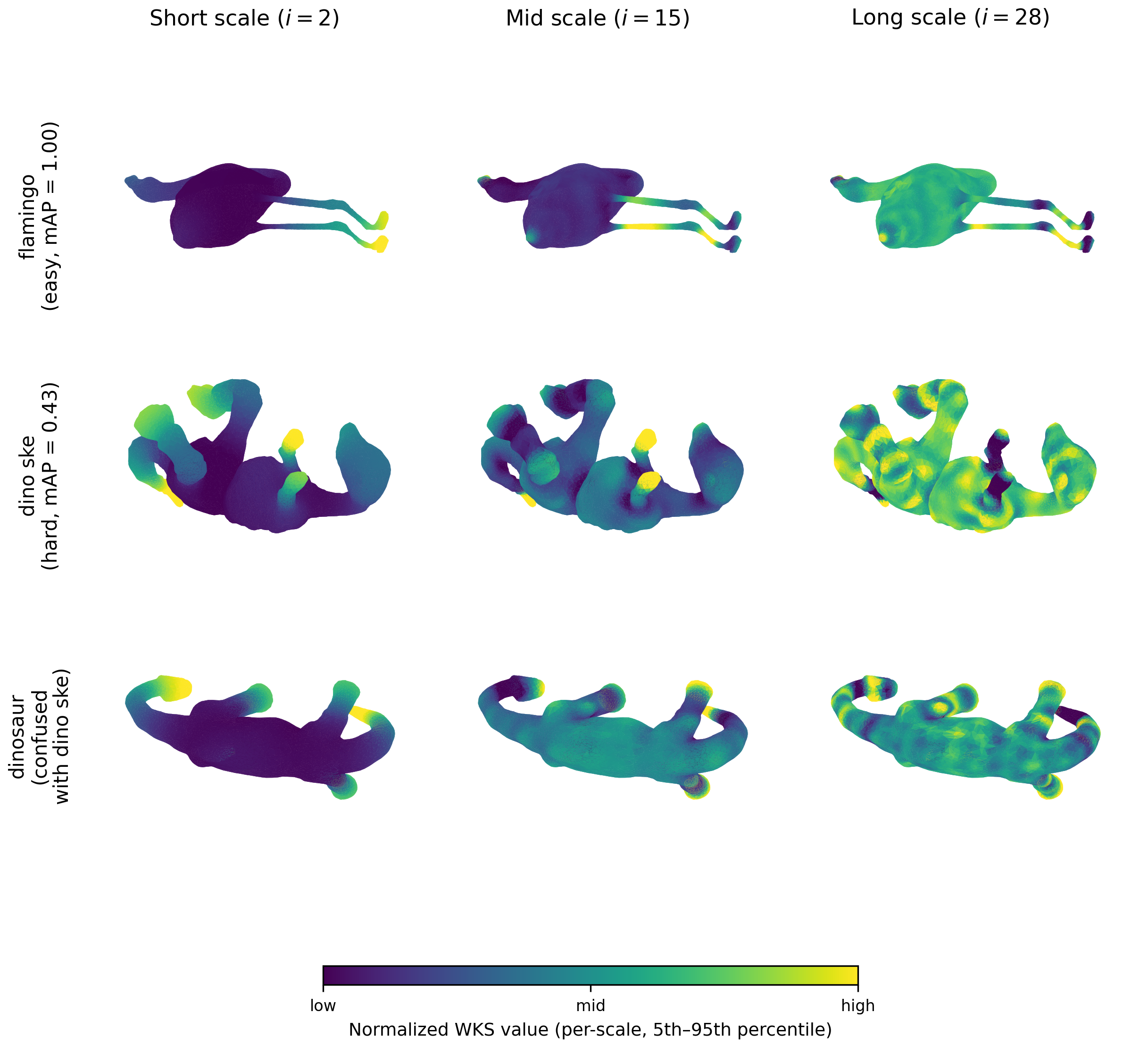}
\caption{Per-vertex WKS responses for three SHREC'11 categories: an easy category (\textit{flamingo}, mAP = 1.00), a hard category (\textit{dino ske}, mAP = 0.43), and a category confused with it (\textit{dinosaur}, confusion rate 0.534). Columns show representative short ($i=2$), mid ($i=15$), and long ($i=28$) scale indices defined in Section~\ref{sec:saliency}. Colors encode per-vertex WKS normalized independently within each scale (5th--95th percentile), so they reflect spatial patterns rather than absolute magnitudes. \textit{Dinosaur} and \textit{dino ske} produce visually related coloring patterns at every scale despite their structural differences, illustrating how WKS can collapse geometrically distinct shapes into overlapping descriptor regions.}
\label{fig:casestudies}
\end{figure}

\subsection{Saliency-Weighted Retrieval}
Saliency-weighted WKS improves overall mAP from 0.810 to 0.867. The improvement is most pronounced on hard categories: mAP rises from 0.517 to 0.673 for the five hardest classes, a gain of 0.156. Figure~\ref{fig:casestudies} visualizes per-vertex WKS responses on representative meshes, where the hard category and its confused counterpart exhibit related spatial patterns that reflect the identified discriminative scale intervals.

\paragraph{Fold stability and sanity check.}
To verify that saliency weighting does not overfit to a single evaluation split, we compute saliency weights independently on three class-balanced folds and apply the resulting weights to the full retrieval set. Folds are constructed by assigning every third shape within each class to one of the three folds, ensuring balanced representation of all 30 categories in each fold. As a sanity check, we also evaluate random nonnegative scale weights averaged over 50 trials. Table~\ref{tab:fold} reports the results. Random weighting fails to reproduce the saliency-weighted gain and performs below the unweighted baseline, confirming that the improvement depends on the identified scale contributions rather than arbitrary reweighting. Cross-fold weighted retrieval closely matches the full saliency result, confirming stability. Furthermore, saliency weighting is negatively correlated with baseline per-class mAP (Spearman $\rho = -0.528$, $p < 0.01$), indicating that gains concentrate on difficult categories.

Saliency-weighted WKS also outperforms the short-only WKS baseline reported in Table~\ref{tab:ablation}, indicating that the method does not simply discard non-short intervals but selectively suppresses harmful scales while retaining useful scale information.

\begin{table}[t]
\caption{Robustness checks for saliency-weighted WKS on SHREC'11.}\label{tab:fold}
\centering
\begin{tabular}{lcc}
\toprule
Method & Overall mAP & Hard-5 mAP \\
\midrule
WKS unweighted & 0.810 & 0.517 \\
WKS random-weighted & 0.804 $\pm$ 0.017 & 0.509 $\pm$ 0.020 \\
WKS cross-fold weighted & 0.865 & 0.667 \\
WKS saliency-weighted & \textbf{0.867} & \textbf{0.673} \\
\bottomrule
\end{tabular}
\end{table}

Table~\ref{tab:perclass} shows per-class mAP for the five hardest and five easiest categories. Saliency weighting provides the largest gains on hard categories such as \textit{ants} and \textit{alien}, while easy categories already near ceiling performance under unweighted WKS. 

\begin{table}[t]
\caption{Per-class mAP for the five hardest and five easiest categories under WKS.}\label{tab:perclass}
\centering
\begin{tabular}{llccc}
\toprule
Group & Class & HKS & WKS & WKS-weighted \\
\midrule
\multirow{5}{*}{Hard}
 & bird2    & 0.559 & 0.395 & 0.521 \\
 & dino ske & 0.364 & 0.431 & 0.588 \\
 & ants     & 0.373 & 0.551 & 0.716 \\
 & pliers   & 0.585 & 0.588 & 0.757 \\
 & alien    & 0.333 & 0.618 & 0.785 \\
\midrule
\multirow{5}{*}{Easy}
 & horse     & 0.829 & 0.992 & 0.988 \\
 & armadillo & 0.724 & 0.997 & 0.990 \\
 & santa     & 0.895 & 0.999 & 0.999 \\
 & two balls & 0.727 & 1.000 & 1.000 \\
 & flamingo  & 0.819 & 1.000 & 1.000 \\
\bottomrule
\end{tabular}
\end{table}

\section{Discussion}

The results reveal two consistent patterns. First, the most useful scale intervals are concentrated rather than uniformly distributed across the scale domain. For HKS, short diffusion times provide strong retrieval signal; for WKS, the most discriminative information concentrates in low energy bands. In both cases, long-scale intervals are actively harmful to retrieval performance. Second, descriptor similarity between class means is a reliable predictor of retrieval failure: pairs with high similarity confuse at rates two orders of magnitude higher than dissimilar pairs. This suggests that the observed retrieval failure is largely structural within the evaluated descriptor space, rather than a consequence of metric choice.

The distinct saliency patterns of HKS and WKS have practical implications for descriptor design. WKS concentrates its discriminative power at low energy scales, which corresponds to smooth, global eigenfunctions, while HKS relies more on mid-range diffusion times. This suggests that frequency-selective descriptors such as WKS are better suited for non-rigid retrieval than heat-diffusion-based ones, and that future descriptors might benefit from explicitly targeting the low-frequency regime.
The random-weight and cross-fold controls indicate that the gain reflects stable scale-importance structure rather than arbitrary dimensional reweighting.

The primary limitation of this study is its reliance on a single benchmark. SHREC'11 contains 20 shapes per class, which limits the statistical reliability of per-class fingerprints. Extending the analysis to larger datasets such as SHREC'15 would strengthen the generalizability of these findings. A second limitation is that our saliency measure is retrieval-level: it quantifies scale importance globally rather than per-shape, which may obscure within-class variation. A third limitation is that mean and max pooling discard spatial distribution of descriptor responses; per-region or attention-based aggregation may reveal finer-grained scale dependence. A fourth limitation is that single-interval masking measures marginal contributions and does not capture joint contributions from spectral coupling, which could be addressed via Shapley-style attribution or pairwise masking.

\section{Conclusion}
We presented a frequency-scale saliency framework for auditing spectral descriptors in 3D shape retrieval. Ablation-based saliency reveals that short scales drive discriminative performance while long scales reduce it, and that HKS and WKS exhibit distinct scale dependence patterns. Class spectral fingerprints show that hard categories lack stable scale structure, and that high descriptor similarity between class pairs is substantially correlated with retrieval failure. Saliency-weighted retrieval improves mAP on hard categories by 0.156, confirming the practical utility of the identified scale intervals. Future work will extend this analysis to larger benchmarks such as SHREC'15 and to learned spectral descriptors and surface networks, comparing scale-saliency patterns between hand-crafted and learned approaches. Cross-fold and random-weight controls further confirm that this improvement reflects stable scale structure rather than arbitrary reweighting.

\begin{credits}
\subsubsection{\discintname}
The author has no competing interests to declare that are relevant to the content of this article.
\end{credits}

%
%
%
\bibliographystyle{splncs04}
\bibliography{references}

@inproceedings{sun2009hks,
  author    = {Sun, J. and Ovsjanikov, M. and Guibas, L.},
  title     = {A concise and provably informative multi-scale signature based on heat diffusion},
  booktitle = {Proceedings of the Symposium on Geometry Processing},
  pages     = {1383--1392},
  year      = {2009}
}

@inproceedings{aubry2011wks,
  author    = {Aubry, M. and Schlickewei, U. and Cremers, D.},
  title     = {The wave kernel signature: A quantum mechanical approach to shape analysis},
  booktitle = {Proceedings of the ICCV Workshops},
  pages     = {1626--1633},
  year      = {2011}
}

@inproceedings{bronstein2010scale,
  author    = {Bronstein, M. M. and Kokkinos, I.},
  title     = {Scale-invariant heat kernel signatures for non-rigid shape recognition},
  booktitle = {Proceedings of CVPR},
  pages     = {1704--1711},
  year      = {2010}
}

@inproceedings{lian2011shrec,
  author    = {Lian, Z. and Godil, A. and Bustos, B. and Daoudi, M. and Hermans, J. and Kawamura, S. and Kurita, Y.},
  title     = {{SHREC}'11 track: Shape retrieval on non-rigid 3{D} watertight meshes},
  booktitle = {Proceedings of the Eurographics Workshop on 3D Object Retrieval},
  pages     = {79--88},
  year      = {2011}
}

@article{biasotti2016recent,
  author  = {Biasotti, S. and Cerri, A. and Bronstein, A. and Bronstein, M.},
  title   = {Recent trends, applications, and perspectives in 3{D} shape similarity assessment},
  journal = {Computer Graphics Forum},
  volume  = {35},
  number  = {6},
  pages   = {87--119},
  year    = {2015}
}

@article{reuter2006laplace,
  author  = {Reuter, M. and Wolter, F. E. and Peinecke, N.},
  title   = {{Laplace-Beltrami} spectra as shape {DNA} of surfaces and solids},
  journal = {Computer-Aided Design},
  volume  = {38},
  number  = {4},
  pages   = {342--366},
  year    = {2006}
}

@article{ovsjanikov2012fmap,
  author  = {Ovsjanikov, M. and Ben-Chen, M. and Solomon, J. and Butscher, A. and Guibas, L.},
  title   = {Functional maps: A flexible representation of maps between shapes},
  journal = {ACM Transactions on Graphics},
  volume  = {31},
  number  = {4},
  pages   = {30},
  year    = {2012}
}

@article{bronstein2011shape,
  author  = {Bronstein, A. M. and Bronstein, M. M. and Guibas, L. J. and Ovsjanikov, M.},
  title   = {Shape {Google}: Geometric words and expressions for invariant shape retrieval},
  journal = {ACM Transactions on Graphics},
  volume  = {30},
  number  = {1},
  pages   = {1},
  year    = {2011}
}

@inproceedings{levy2006laplace,
  author    = {L{\'e}vy, B.},
  title     = {{Laplace-Beltrami} eigenfunctions towards an algorithm that understands geometry},
  booktitle = {Proceedings of the IEEE International Conference on Shape Modeling and Applications},
  pages     = {13},
  year      = {2006}
}

@inproceedings{rustamov2007laplace,
  author    = {Rustamov, R. M.},
  title     = {{Laplace-Beltrami} eigenfunctions for deformation invariant shape representation},
  booktitle = {Proceedings of the Symposium on Geometry Processing},
  pages     = {225--233},
  year      = {2007}
}

@article{reuter2009discrete,
  author  = {Reuter, M. and Biasotti, S. and Giorgi, D. and Patan{\`e}, G. and Spagnuolo, M.},
  title   = {Discrete {Laplace-Beltrami} operators for shape analysis and segmentation},
  journal = {Computers and Graphics},
  volume  = {33},
  number  = {3},
  pages   = {381--390},
  year    = {2009}
}

@inproceedings{kokkinos2012intrinsic,
  author    = {Kokkinos, I. and Bronstein, M. M. and Litman, R. and Bronstein, A. M.},
  title     = {Intrinsic shape context descriptors for deformable shapes},
  booktitle = {Proceedings of CVPR},
  pages     = {159--166},
  year      = {2012}
}

@article{litman2014learning,
  author  = {Litman, R. and Bronstein, A. M.},
  title   = {Learning spectral descriptors for deformable shape correspondence},
  journal = {IEEE Transactions on Pattern Analysis and Machine Intelligence},
  volume  = {36},
  number  = {1},
  pages   = {171--180},
  year    = {2013}
}

@book{bronstein2008numerical,
  author    = {Bronstein, A. M. and Bronstein, M. M. and Kimmel, R.},
  title     = {Numerical Geometry of Non-Rigid Shapes},
  publisher = {Springer},
  address   = {New York},
  year      = {2009}
}

@article{rodola2017partial,
  author  = {Rodol{\`a}, E. and Cosmo, L. and Bronstein, M. M. and Torsello, A. and Cremers, D.},
  title   = {Partial functional correspondence},
  journal = {Computer Graphics Forum},
  volume  = {36},
  number  = {1},
  pages   = {222--236},
  year    = {2017}
}

@inproceedings{tombari2010unique,
  author    = {Tombari, F. and Salti, S. and Di Stefano, L.},
  title     = {Unique signatures of histograms for local surface description},
  booktitle = {Proceedings of ECCV},
  pages     = {356--369},
  year      = {2010}
}

@article{wan2017full,
  author  = {Wan, L. and Zou, C. and Zhang, H.},
  title   = {Full and partial shape similarity through sparse descriptor reconstruction},
  journal = {The Visual Computer},
  volume  = {33},
  pages   = {1497--1509},
  year    = {2017}
}

@article{pickup2018evaluation,
  author  = {Pickup, D. and Liu, J. and Sun, X. and Rosin, P. L. and Martin, R. R. and Cheng, Z. and Lian, Z. and Nie, S. and Jin, L. and Shamai, G. and Sahillio{\u{g}}lu, Y. and Kavan, L.},
  title   = {An evaluation of canonical forms for non-rigid 3{D} shape retrieval},
  journal = {Graphical Models},
  volume  = {97},
  pages   = {17--29},
  year    = {2018}
}

\end{document}